\documentclass[a4paper]{jpconf}
\usepackage{graphicx}
\usepackage{xcolor}

\begin{document}
\title{Boosting I/O and visualization for exascale era using Hercule: test case on RAMSES}

\author{Loic Strafella and Damien Chapon}

\address{CEA/IRFU/LILAS Centre d'Études de Saclay, 91191 Gif-Sur-Yvette, France} 

\ead{loic.strafella@cea.fr, damien.chapon@cea.fr}

\begin{abstract}
It has been clearly identified that I/O is one of the bottleneck to extend application for the exascale era. New concepts such as 'in transit' and 'in situ' visualization and analysis have been identified as key technologies to circumvent this particular issue. A new parallel I/O and data management library called Hercule, developed at CEA-DAM, has been integrated to Ramses, an AMR simulation code for self-gravitating fluids. Splitting the original Ramses output format in Hercule database formats dedicated to either checkpoints/restarts (HProt format) or post-processing (HDep format) not only improved I/O performance and scalability of the Ramses code but also introduced much more flexibility in the simulation outputs to help astrophysicists prepare their DMP (Data Management Plan). Furthermore, the very lightweight and purpose-specific post-processing format (HDep) will significantly improve the overall performance of analysis and visualization tools such as PyMSES 5. An introduction to the Hercule  parallel I/O library as well as I/O benchmark results will be discussed.
\end{abstract}

\section{Introduction}

The race to exascale leads to many-core architectures on which the gap between computational power and I/O bandwidth keeps increasing. On such many-core supercomputers, classical sequential one file-per-process I/O paradigm leads to bottlenecks that needs to be removed in order to keep scaling scientific code. In the meantime, scientists develop more complex and multi-scale models for an higher fidelity simulations that will produce tremendous amount of data to manage and post-process. There is a real need for parallel I/O with sustainable performance as well as data management libraries. Different solutions exist such as MPI-IO, Parallel HDF5, Parallel NetCDF among which only HDF5 allow the user to organize its data. In this work, we focused on the integration of Hercule \cite{Hercule}, a library developed at CEA-DAM which was designed for efficient parallel I/O and for data management. As a test case scenario we used an astrophysical code called Ramses \cite{Ramses} which present I/O scalability issues. Previous work on Ramses I/O by Wautelet and Kestener \cite{Wautelet} concluded that the complex I/O pattern induced by adaptive mesh refinement (AMR) data structure in Ramses makes POSIX I/O more efficient than parallel I/O libraries. Nonetheless, with a fine tuning and a high number of cores MPI-I/O performances become comparable to POSIX one file-per-process approach but was not able to improve I/O efficiency or scalability.\\

In addition to one file-per-process paradigm, the last available version of Ramses uses only one kind of output data flow. Thus the output format mixes the requirements for both checkpoint/restart and post-processing purposes. On the top of that, physical data in Ramses I/O routines are ordered level by level according to the AMR mesh structure of the current MPI process. Inside the same level all physical fields are written one after the other, which means that all the data are nested in the binary file and cannot be accessed without reading the entire file according to the AMR structure. In addition, some pre-processing is done in order to compute extra physical quantities that could be used by post-processing tools. This pattern is used by all modules of Ramses: hydrodynamics (HYDRO), gravity solver (POISSON), magnetohydrodynamics (MHD), radiative transfer (RT). Some other additional quantities such as particles are written in a more classical fashion. Even if that pattern and the separation of quantities in multiple files seems advantageous for post-processing, it leads to unnecessary overhead for both use cases. Such an output machinery leads to several issues such as file systems saturation by the number of files, many data redundancy, heavy post-processing steps and a very difficult data management plan.\\

In this work, we first discuss how we change the I/O pattern of Ramses with the integration of Hercule library by introducing its new database format: \textit{HProt} and \textit{HDep}. In a second time, we discuss how we successfuly reduced by a factor 2 the post-processing data volume by removing redundancy in AMR tree structure by adding lossless data compression techniques for AMR tree structure and double precision data to produce a very lightweight post-processing format. Thirdly, we discuss benchmarks results that show a significant improvement of I/O scalability and finally, we put the emphasis on recent AMR visualization techniques developped by the CEA-DAM and already available on VTK package \cite{Harel}, \cite{Lekien}, that can be interfaced with Hercule format in order to make a pipeline with a data centric approach.

\section{Integration of Hercule library} 
Hercule one file-for-multiple-processes approach means that a single file is shared between multiple MPI processes called \textit{contributors}. Thus a simulation ran with N MPI processes and P number of contributors per file (NCF) will create at least N/P files. Indeed, all the data of multiple time steps from the different contributors are stored in the same file until a user defined file size limit is reached and a new file is created. Because of this data aggregation the number of files created by a simulation can be significantly reduced from tens of thousand by Ramses legacy to only hundreds of files. In addition, a coherent data structure is defined by the resulting files and is called an \textit{Hercule database} in which a \textit{context} is all the data of a specific time step and a \textit{domain} is all the data of a specific contributor in a context. There is two types of \textit{databases} used in this work: \textit{HProt} for checkpoint/restart and \textit{HDep} for post-processing. A free format storage with basic features is provided by HProt database which are only used by the code itself as checkpoint. More advanced features with data model are provided by HDep database which was designed to be generic in order to be shared by other codes such as post-processing and visualization tools.\\

\begin{figure}[!ht]
\begin{center}
\includegraphics[width=26pc]{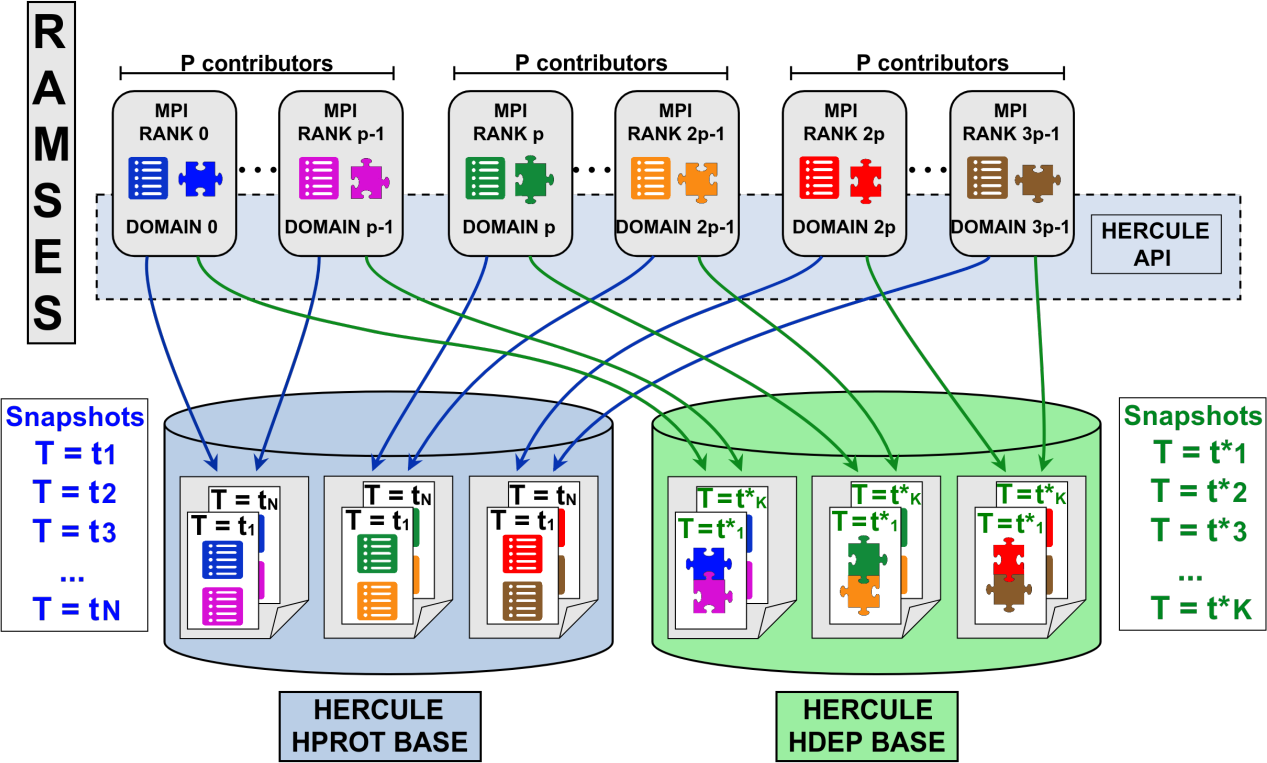}
\end{center}
\caption{\label{RamsesHercule} Schematic data flow of Ramses with Hercule library filling databases for checkpoint/restart and post-processing with a different output frequency.}
\end{figure}

Based on those two kind of databases, the original data flow of Ramses was split in order to separate checkpoint from post-processing output thus introducing more flexibility and opportunity to optimize each data flow. Indeed, as shown in figure \ref{RamsesHercule}, the new version of Ramses with Hercule produces two databases, each of which are fed with data from groups of contributors with different frequencies. In addition, since HProt databases are not meant to be analyzed or shared by different code, only raw data are sent by contributors. On the opposite, HDep databases will be used by other codes thus contributors build standardized data object according to an Hercule AMR model that we discuss later. \\

Different approaches with different levels of granularity of data aggregation were tested in order to optimize checkpoint/restart data flow. We started out by following the original pattern and replaced the fine data granularity of Ramses I/O calls by Hercule calls. But the huge number of I/O calls with small size of data block induced significant library overhead that lead to very poor I/O efficiency. We completely changed the I/O pattern in order to increase the data granularity of our I/O strategy to reduce the number of I/O calls and to make big blocks of untransformed raw data. With this second approach, the use of preallocated array for data manipulation prior to checkpoint outputs and after restart inputs is no longer needed. By removing array allocation and data manipulation we obtained slight acceleration which is exactly compensated by library overheads. This leads to I/O performances similar to Ramses legacy below 2048 MPI processes.

\begin{figure}[h!]
\centering
\includegraphics[width=36pc]{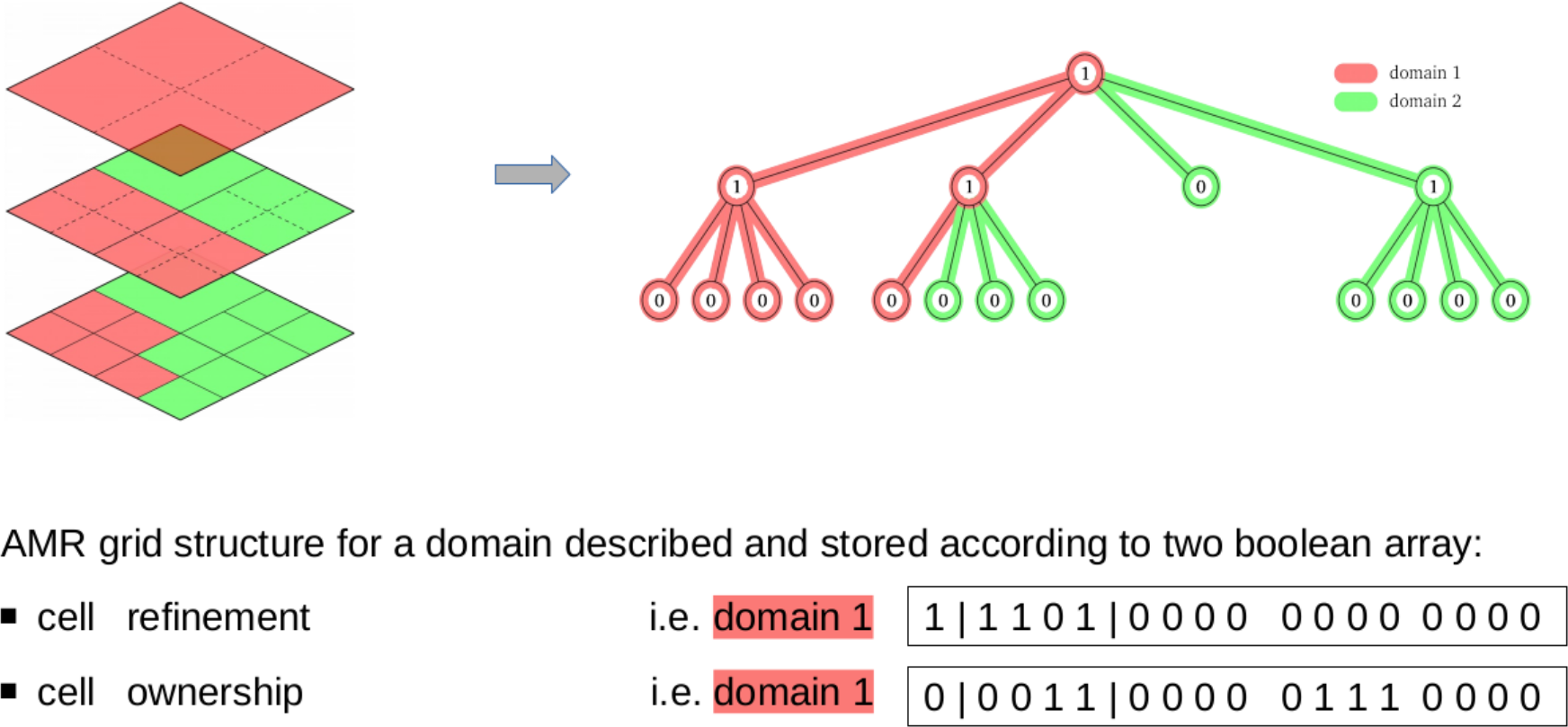}
\caption{\label{TreeStructure} 2D mesh structure split over two process and its tree-based representation. Red color shows mesh elements that belong to process 1 and green color those belonging to process 2. Cell refinement and cell ownership arrays are shown for process 1 according to Hercule AMR model.}
\end{figure}

HDep offer standardized, optimized and self-describing data model for different kind of meshes structure, in this work, the AMR-3D data model was used. Using this model, a data object representing  a piece of the AMR structure and its associated physical quantities is built by each MPI process and stored in the database. Each object is self-describing thus any post-processing tool that uses Hercule API is able to assemble the objects in order to built the entire AMR tree structure with its physical data. To achieve a self-describing structure each object has a number of attributes which define an AMR tree. The most important ones are the refinement array and the ownership array which are used to describe the mesh structure. Indeed, since AMR structure can be seen has a tree-based structure, Hercule use boolean value to describe the tree according to the binary state of a cell: coarse or leaf. According to a defined path which goes from the top to the bottom level by level from left to right, an array of boolean values for each node is built, as shown in figure \ref{TreeStructure}. Of course, physical quantities are added to the object by following the same strategy. To provide more flexibility to the user and to reduce the amount of data, we provided to the user the ability to select a subset of physical quantities to output in the HDep database through the Ramses configuration input file. The ownership array adds a complementary information about cells and their belonging to the current MPI process. This extra array is used for post-processing purposes to assemble different parts of the tree by taking account of ghost cells. Based on those two arrays we developed a tree pruning algorithm in order to reduce the important redundancy in the data.

\subsection{Tree pruning algorithm to remove redundancy}

Because of the Hilbert space filing curve, domain boundaries of Ramses can occur on leafs of the tree and at different levels. This method was chosen in order to achieve a good balancing of mesh elements between MPI processes which is an important process mainly due to the AMR technique which creates and destroys dynamically mesh elements during the simulation. On the top of that, the multigrid solver of Ramses requires that each MPI process had a global but degraded information of the mesh structure of the entire simulation box. This extra information required by the solver can be removed from the post-processing data because it makes the AMR tree unnecessary heavy and redundant over all processes. Therefore, we developed an efficient tree pruning algorithm to reduce the AMR tree described by a process thus reducing the number of cells in the tree of the current process and consequently the number of associated physical quantities.\\

The tree pruning algorithm analyzes the tree-based structure from the bottom to the top and dynamically changes the refinement values of unnecessary cells which are defined as ghost coarse cells of whom leafs are also all ghosts. We used the data of a simulation project that treats the evolution of molecular clouds under self-gravity and MHD turbulence \cite{Orion} called Orion as a test case. We noticed that the pruning algorithm was able to achieve an average reduction on all MPI processes of about 31.3 \% with a worst case scenario of about 17.2\% and a best case scenario of about 47.3\%, which means that there is a lot of redundancy in the data, see figure 3 for a percentage of removed cells by domain. Those results makes the tree pruning the most important process in order to reduce the data volume of the post-processing database (HDep) with this AMR based code. 

\subsection{Lossless compression of AMR data}

\begin{figure}[h!]
\centering
\begin{minipage}[t]{14pc}
\includegraphics[width=16pc]{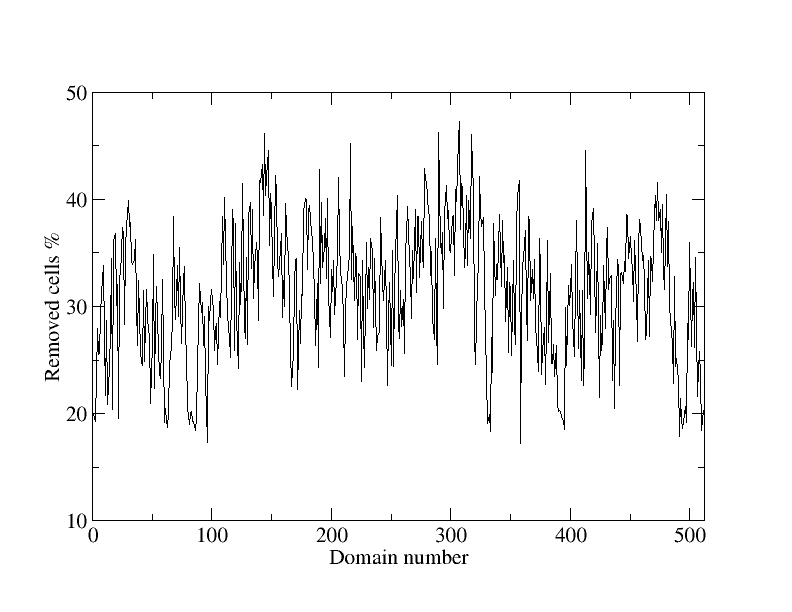}
\caption{\label{TreePruning}Percentage of removed cells by domain (MPI process) for Orion data using the implemented tree pruning algorithm.}
\end{minipage}\hspace{4pc}%
\begin{minipage}[t]{14pc}
\includegraphics[width=16pc]{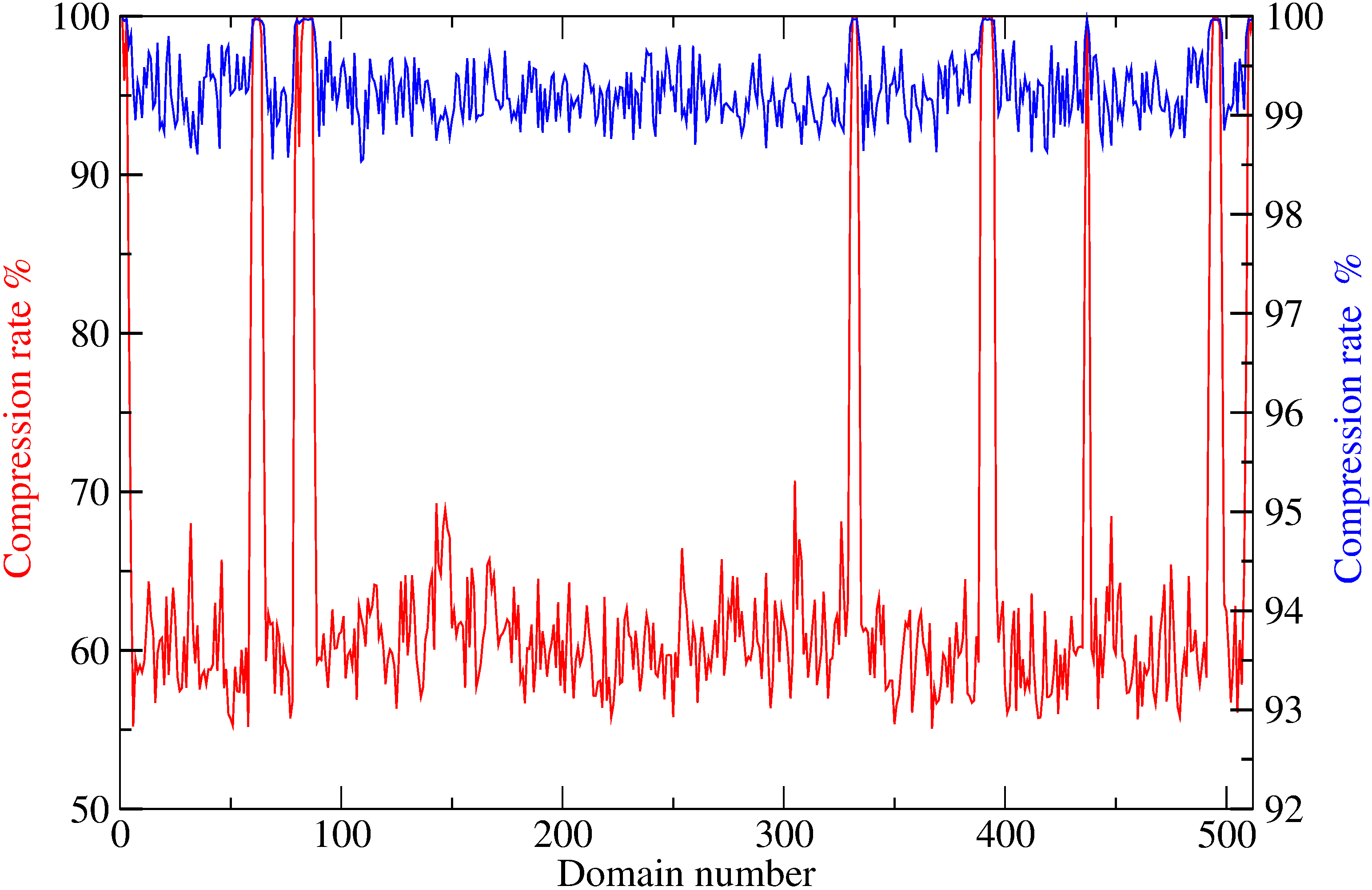}
\caption{\label{CompressionAMRGhost}Compression rate by domain (MPI process) obtained using the base-52 and character encoding algorithm on refinement in red and ownership array in blue, for Orion data.}
\end{minipage} 
\end{figure}

Once the tree is defined and reduced through the pruning algorithm, we developed compression algorithm for the booleans arrays that defined the AMR tree structure. Indeed, even if those arrays are boolean arrays, one byte is still used to store only two values namely true or false. Without introducing data compression, the most compact way to store such arrays would be to use a single bit and a bit field structure for those arrays. We proposed a even more compact way to store those arrays by using a fast lossless compression algorithm that use base-52 and character encoding in order to significantly reduce the size of those arrays. By taking a random MPI process that contains about one million of cells we compared the achieved compression with a bit field equivalent. The refinement array would take 0.12 MB on a bit field structure while with the compression method it would only require a string of 1.5 KB with a compression time of only 0.5 ms. In the figure \ref{CompressionAMRGhost}, we show in red the compression ratio between the resulted string from the compression algorithm and a bitfield equivalent for the refinement array and in blue for the ownership array. On average we achieved 63.4\% of compression on the refinement array and 99.3\% on the ownership array. Such huge compression rates are explained by the fact that those array contains a lots of consecutive zeros and ones. Nonetheless, the effectiveness of the algorithm depends on the chaoticity of the AMR mesh. Indeed, the compression of the ownership array is better because the number of consecutive zeros or ones has less fluctuations. At this point, the volume of data used to describe the AMR structure of the simulation becomes a negligible part of the overall data volume.

\subsection{Lossless compression of double precision data}
Because Hercule library does not provide any compression functionality we developed a fast and customizable at runtime delta compression technique which takes advantage of the AMR tree structure itself as well as the storage order of the tree structure in the refinement array. Usually, in delta compression an extrapolation function is used in order to predict a value from which the delta is computed and stored. The idea behind this algorithm called father-son prediction is to take advantage of the inner hierarchy of an AMR mesh as a way to predict a value. Indeed, in the Ramses code a coarse cell called the father has its own value which depends on the value of its leafs cells, called sons. We proposed to use the value of a father cell as a predictor value for its sons'. The prediction as well as son values are mapped on 64 bits integers in the case of 64 bits floats, then an XOR operation is done between the prediction and all son values to obtain a list of residues. An other XOR operation is done between all computed residues and the maximum number of leading zeros that could be removed from the binary representation without any loss of precision is computed. Finally, a bitfield is built from the remaining bits of each residue in order to obtain the final compressed field.\\

A dedicated format is used for the compression field which is based on the encoding of the number of leading zeros which have been removed from the residue. As a default value, we chose to use 4 bits to encode the number of removed zeros which allow to remove 15 zeros maximum. Nevertheless, this parameter can be optimized at runtime in order to remove more zeros in the case, for example, of physical fields which only vary locally. With a default value of 4 bits the maximum asymptotic compression rate is 22.65\%. Because of the father-son relationship,s the compression algorithm goes from the top of the AMR tree to the bottom which means that the decompression process could be done only partially. For example, visualization tools could load and uncompress data down to a defined level of refinement thus saving memory which is often a bottleneck in the visualization process. On the other hand, it also means that the user cannot access access a specific value without decompressing all the father cells. Finally, since the father value is used as predictor, this compression technique works well for non conservative physical quantities. If one wants to use it for conservative quantities, a multiplicative factor has to be applied to the father value according to the refinement strategy.\\

\begin{figure}[h!]
\centering
\begin{minipage}[t]{14pc}
\includegraphics[width=16pc]{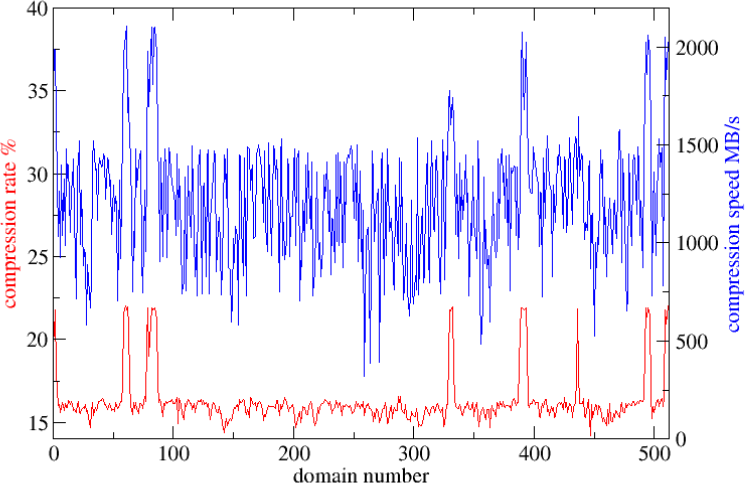}
\caption{\label{CompressionDataDensity}Compression rate in red and compression speed in blue as a function of the domain number for the density field in Orion data. Average values for compression rate is 16.26\% and for compression speed is 1321.40 MB/s.}
\end{minipage}\hspace{4pc}%
\begin{minipage}[t]{14pc}
\includegraphics[width=16pc]{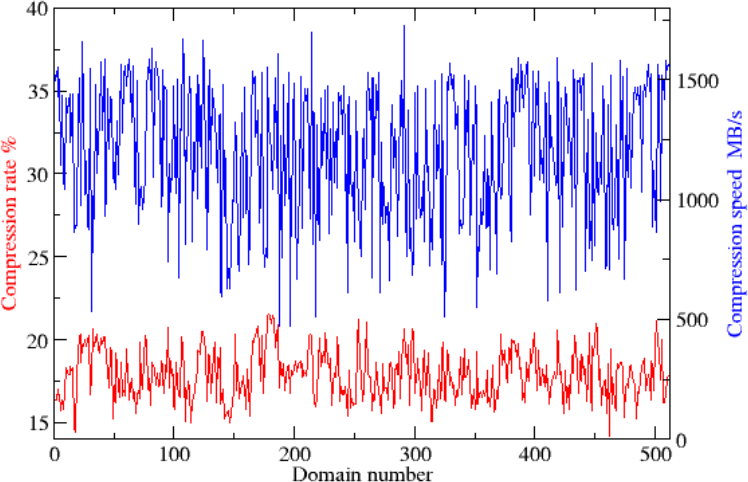}
\caption{\label{CompressionDataVy}Compression rate in red and compression speed in blue as a function of the domain number for the velocity "y" field in Orion data. Average values for compression rate is 17.91\% and for compression speed is 1285.90 MB/s.}
\end{minipage} 
\end{figure}

In order to benchmark this algorithm we used the Orion data \cite{Orion}. As shown in the figure \ref{CompressionDataDensity} for the density field and in figure \ref{CompressionDataVy} for the $y$ component of the velocity vector, we were able to achieve a good compression speed using a sequential version of the compression algorithm on an Intel i5 @ 2.5Ghz, 7th gen, 3Mo cache processor. On average, we removed 11 leading zeros on sons' values for the density field and 12 zeros for the $y$ component of the velocity field which means that we removed the most of the exponent bits. We developed a sequential version of the (de)compression algorithm but it could be trivially parallelized/vectorized using multiple seed of father cells values.

\section{Benchmarks analysis}
\begin{figure}[!ht]
	\centering
	\includegraphics[width=26pc]{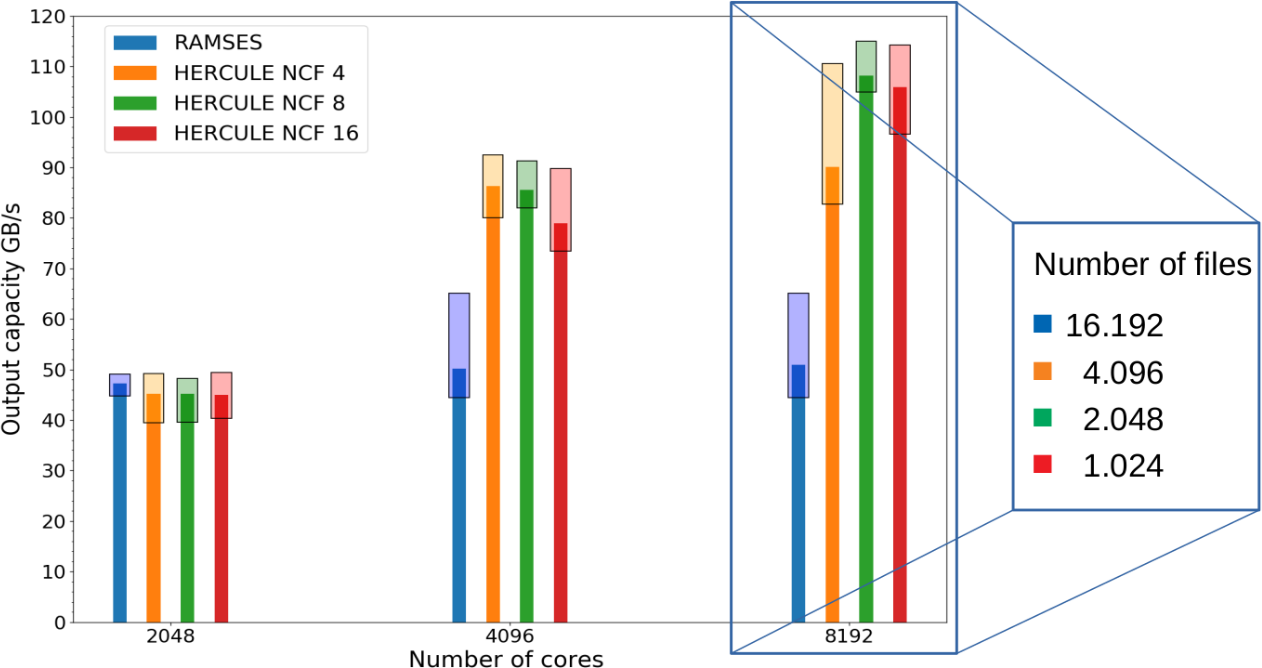}
	\caption{\label{IObench} Strong scaling of writing throughput of Sedov3D with Ramses Posix I/O versus Ramses with Hercule library with different numbers of contributors per file and fixed stripe size. On the right, the number of files generated by one benchmark at 8192 MPI processes with Ramses and different value of the NCF parameter with Ramses-Hercule.}
\end{figure}

In this section, we discuss strong scalability benchmarks ran on the TGCC Joliot-Curie supercomputer on the scratch partition which throughput capacity was evaluated at 300 GB/s. The benchmarks were ran in a user like environment with no specific privileges and during a normal load of the supercomputer which means that other users may have done I/O operations at the same time that can explain high standard deviation in our results. We used a Ramses 3D test case: Sedov3D (explosion in a a cubic box) using a domain size of $2048$ cells in each direction of space. AMR and time integration were deactivated during the execution thus the load was perfectly balanced between all the MPI processes. We measured the output capacity of Ramses namely the amount of gigabyte written by the code by unit of time. We increased the number of MPI processes from 2048 to 8192 while keeping the size of the problem constant. All benchmarks were ran 5 times and we present the average I/O bandwidth as well as the standard deviation, see figure \ref{IObench}. \\

Ramses legacy benchmarks generated two series of file per MPI processes according to the one-file-per-process paradigm: an AMR file with a complex pattern and a much heavier HYDRO file with 5 scalars fields. No specific I/O parameters were set for those legacy benchmarks. On the other hand, Ramses with Hercule benchmarks required to set 4 parameters: the maximum file size, number of contributor per file, the stripe count and the stripe size. The two first parameters are Hercule parameters and we fixed the maximum file size to 2 GB which is its default value while we ranged the second parameter from 4 to 16. The last two parameters are Lustre file system parameters and can be set by the command \textit{lfs setstripe}. After a preliminary parametric study of the impact of the stripe size on I/O performance, we fixed a stripe size of 4 MB because I/O performance were much better than the default value of 1 MB. A bigger stripe size can be set until a maximum value of 4 GB according to Lustre documentation depending on the file size but in our case the installed Lustre file system was not parameterized to take advantage of a stripe size bigger that 4 MB. We did an equivalent study for the stripe count which is the number of OST on which the file is stripped and we came to the conclusion that setting setting the stripe count at the number of contributors per file leads to optimal performances.\\

The blue bars refer to Ramses legacy performances while the orange, green and red ones refer to Hercule performances with different values for the number of contributors. Firstly, we noticed that the performance of Ramses POSIX and Ramses-Hercule are similar at 2048 MPI processes which means that the parallel I/O library show good performance in I/O bandwidth despite an unavoidable overhead in communication induced by parallel I/O. When increasing the number of MPI processes to 4096, Ramses legacy does not scale and show similar performance than the 2048 benchmark. At the opposite, the Hercule version of Ramses was able to scale the I/O bandwidth for the 3 values of the NCF parameter. We almost achieved a factor 2 in I/O performance. Increasing the number of MPI processes to 8192 clearly shows a saturation of Ramses legacy I/O performances while we were able to keep improving the performances with Hercule. We did not achieved a factor 2 in performance probably because the amount of data by process with a problem size of $2048^3$ is too small. In order, to keep increasing performance I/O performance, one needs to increase the amount of data per MPI process and increase the default value for the maximum file size of Hercule. Nevertheless, the very good performance obtained at 8192 processes with a number of contributors of 16 MPI processes per file shows that we successfully improved the I/O performance of Ramses using the Hercule library. Finally, the number of files generated per benchmark were significantly reduced. As shown on the table in figure \ref{IObench} for 8192 processes, we successfully reduced the number of files by 16 while improving the I/O bandwidth by a factor of 2.2. 

\section{Visualization and future work}
\begin{figure}[!ht]
	\centering
	\includegraphics[width=18pc]{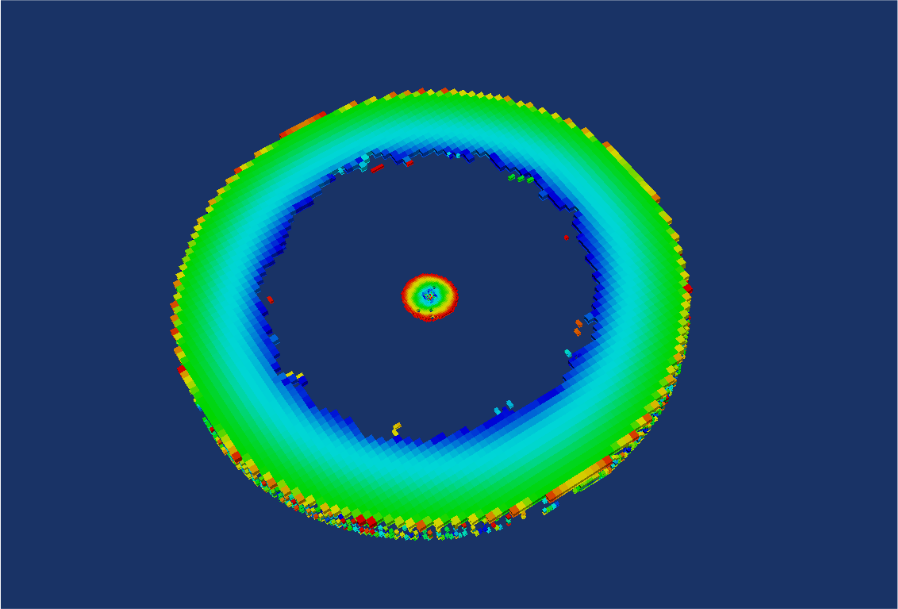}
	\caption{\label{Galax}Visualization of a small test case of a galaxy using the VTK HyperTreeGrid structure obtained with a custom version of Pymses that use Hercule library. Two threshold filters have been applied in order to show the central region which contains high density materials and the outer region with lower density materials. AMR cells are displayed.}
\end{figure}

At the DIF departement of the CEA, a parallel visualization tool, Love, based on VTK and Paraview was developed in order to post-process and visualize large amount of data of AMR-based simulation code, \cite{Love}. One of the goal of Love was to provide support on VTK for generic AMR tree-based data sets thus an \textit{HyperTreeGrid} data class as well as dedicated filters was integrated to the VTK package. Based on the Hercule post-processing HDep database and its AMR data model we can to take advantage of those new available development to visualize AMR data produced with our version of Ramses-Hercule. We are currently working on the integration of Hercule and the VTK library to Pymses, a post-processing and visualization tool dedicated to Ramses outputs developed at CEA-Irfu. As a first results, we successfully visualized a HDep data produced by a small test case, see figure \ref{Galax}. We show the result of two threshold filters applied on the density field. One for high values of density which result in the inner circle and one for small values of density that leads to the outer circle.  

\section{Conclusion}
We successfully integrated Hercule, a parallel I/O and data management library, in the astrophysical code RAMSES, that presented I/O scalability issues. We changed the original checkpoint/restart file format of RAMSES and introduced a new output data flow for post-processing. Thus the new version of RAMSES with Hercule produces HProt databases, dedicated to checkpoint/restart, and HDep databases, dedicated to post-processing. We developed an AMR tree pruning algorithm that significantly reduce the redundancy found in RAMSES AMR data which leads to a significant reduction of the data volume required for post-processing, as shown with the Orion data, \cite{Orion}. By adding lossless data compression for the AMR tree structure to the HDep AMR data model, the description of the AMR structure is reduced to a negligible part of the overall data. On the top of that, we also developed a fast lossless delta compression algorithm for AMR data which use the inner hierarchy of AMR in order to compress non conservative physical fields. \\

Benchmark results between RAMSES legacy and RAMSES-Hercule show a significant improvement of the I/O bandwidth as well as an improved I/O scalability of RAMSES at 8192 MPI processes with Hercule. Finally, we integrated the Hercule and VTK library in Pymses, a dedicated post-processing tools, in order to take advantage of new visualization techniques for AMR data setss, open sourced by the CEA and available on VTK library. We used the HyperTreeGrid class and the HyperTreeGrid threshold filter to produce the first image of an HDep database produced by the new version of RAMSES-Hercule. In the near future, new specific filters for AMR data sets such as volume rendering will be developed as well as pipelines between simulation code and post-processing tools in order to achieve real time visualization.

\ack{This work was performed using HPC resources from GENCI-TGCC (Grant 2019-A0050402192)}

\end{document}